\documentclass[11pt,leqno]{article}

\usepackage[top=3cm, bottom=3cm, left=4cm, right=4cm ]{geometry}

\usepackage{amsmath,amssymb,amsthm}
\usepackage{enumitem}

\begin{document}

\title{ Boundary Conditions at Infinity for Physical Theories\footnote{This is an author-created version of an article presented by L. Infeld on April 12, 1958 to and published in Bulletin de l'Acad\'emie Polonaise des Sciences, S\'erie des sci. math., astr. et phys. Vol. {\bf VI}, No. 6 (1958) 403-406.}}
\author{by\medskip\\Andrzej Trautman}
\date{\empty}
\maketitle

\paragraph{1.} The Cauchy problem is the most natural for hyperbolic partial 
differential equations. When dealing with physical problems, we are, 
however, often interested in solutions of field equations with given sources 
when nothing is known about initial conditions. A whole set of fields 
corresponds, in general, to given sources and, in order to arrive at a unique 
solution of the problem, we must specify some additional condition. 
For linear field equations this condition may consist in prescribing the 
form of Green's function (e.g. retarded, advanced, etc.). If we investigate 
the field in the whole (unbounded) space-time we can ensure uniqueness 
by specifying some appropriate {\em boundary conditions} at spatial infinity. 
The latter approach has the advantage of being applicable to non-linear 
theories, such as the theory of general relativity. These boundary conditions, first formulated for a periodic scalar field by Sommerfeld \cite{somm}, 
have a definite physical meaning. E. g., the ``Ausstrahlungsbedingung'' 
of Sommerfeld means that the system can lose its energy in the form 
of radiation and that no waves are falling on the system from the exterior. The purpose of this paper is to formulate boundary conditions for 
scalar and Maxwell theories in a form which exhibits their physical 
meaning and is proper to a generalization for the gravitational case.

\paragraph{2.} Let us first take the {\em scalar wave equation}
\begin{equation}
\Delta\varphi-\frac{\partial^2\varphi}{\partial t^2}=-4\pi\rho
\label{wave-eq}
\end{equation}
and assume $\rho(\vec{r},t)$  to be a regular function vanishing outside a bounded 3-dimen\-sio\-nal region $V$. The retarded solution of \eqref{wave-eq} can be written in 
the form
\begin{equation}
\varphi(\vec{r},t)=\int_V\frac{\rho(\vec{r}{\,}^{\prime},t-R)}{R}\,dV', \quad\quad R=|\vec{r}-\vec{r}{\,}^{\prime}|.
\label{ret-sol}
\end{equation}
From \eqref{ret-sol} we obtain the following  asymptotic values of $\varphi$ and its derivatives\footnote{$\Phi_{A}=O(r^{-k})$ means that there exists a constant $M$ such that, for a sufficiently large $r$, we have $|\Phi_A|<Mr^{-k}$; Greek indices run from 0 to 3, Latin ones---from 1 to 3; $x^{0} = t$, $(x^1, x^2, x^3) = \vec{r}$; a comma followed by an index denotes differentiation. Summation convention will be used throughout. Indices will be raised by means of the Galilean metric tensor $\eta^{\alpha\beta}$  ($\eta^{00} = 1$, $\eta^{i0} = 0$, $\eta^{ik}= -\delta^{ik}$). Square brackets stand for alternation, e. g., $\Phi_{[\mu\nu\rho]}=\Phi_{\mu\nu\rho}+\Phi_{\nu\rho\mu}+\Phi_{\rho\mu\nu}-\Phi_{\rho\nu\mu}-\Phi_{\nu\mu\rho}-\Phi_{\mu\rho\nu}$.}:
\begin{equation}
\begin{gathered}
  \varphi=r^{-1}\int_V\rho(\vec{r}{\,}^{\prime},t-R)\,dV'+O(r^{-2}),\\
  \varphi_{,\sigma}=k_{\sigma}r^{-1}\int_V\rho_{,\sigma}(\vec{r}{\,}^{\prime},t-R)\,dV'+O(r^{-2}),
\end{gathered}
\label{phiphi,}
\end{equation}
where
\begin{equation}
k^\sigma=(1,n^s),\quad\quad n^s=x^s/r
\label{k}
\end{equation}
is a null vector field.

Now, we can formulate the following boundary conditions to be imposed on solutions of \eqref{wave-eq}:
\begin{enumerate}[label=(\arabic*),leftmargin=0.7cm]
\setcounter{enumi}{\value{equation}}
\item $\varphi=O(r^{-1})$ \label{phi-O}; 
\item {\em there exists a function $\psi=O(r^{-1})$ such that $\varphi_{,\nu}=\psi k_\nu+O(r^{-2})$, where $k_\nu$ is given by  \eqref{k}.} \label{psi}
\end{enumerate}
\setcounter{equation}{\value{enumi}}

We see from Eqs. \eqref{phiphi,} that every retarded solution of \eqref{wave-eq} fulfills \ref{phi-O} and \ref{psi}. Conversely, if condition \ref{psi} is fulfilled, then $\varphi$  satisfies Sommerfeld's radiation condition
\[
\lim_{r\to\infty}rk^\nu\varphi_{,\nu}=\lim_{r\to\infty} r(\partial\varphi/\partial t+\partial\varphi/\partial r)=0.
\]

Thus, the wave equation with a spatially bounded source has always one, and only one, solution fulfilling our conditions \ref{phi-O} and \ref{psi}.

If we replace \eqref{k} by $k^\sigma=(1,-n^s)$  we obtain the conditions which characterize advanced solutions of \eqref{wave-eq}.

Let us introduce the energy-momentum density tensor of the field $\varphi$:
\[
T^\nu{}_\mu=L\delta^\nu{}_\mu-\varphi_{,\mu}\partial L/\partial\varphi_{,\nu},\quad\quad\text{where}\quad\quad L=-\eta^{\rho\sigma}\varphi_{,\rho}\varphi_{,\sigma}/8\pi.
\]
Taking into account the asymptotic expressions of $\varphi_{,\nu}$  we have $L = O(r^{-3})$ 
and
\begin{equation}
4\pi T_{\mu\nu}=\psi^2k_\mu k_\nu+O(r^{-3}).
\label{T}
\end{equation}

The asymptotic form of $T_{\mu\nu}$  resembles the energy-momentum tensor 
of a perfect fluid with vanishing rest mass. Further, we can obtain from \eqref{T}
the time rate of radiated energy and momentum:
\[
4\pi W_\mu=4 \pi \oint_ST^s{}_\mu n^s \,dS=\oint_S\psi^2k_\mu\,dS.
\]

The integrals are to be taken over the surface of a sphere ``at infinity''. The condition \ref{psi} with Eq. \eqref{k} ensures that $W_0\geq 0$.

\paragraph{3.} The situation is somewhat more complicated in {\em electrodynamics} 
because of the gauge-invariance. Maxwell's equations
\begin{align}
f^{\mu\nu}{}_{,\nu}&=-4\pi j^\mu, & f_{\mu\nu}&=A_{[\nu,\mu]}\label{maxw}
\end{align}
can be reduced to four wave equations
\begin{equation}
\Delta A^\mu-\frac{\partial^2 A^\mu}{\partial t^2}=-4\pi j^\mu
\label{em-wave}
\end{equation}
if one imposes on potentials the Lorentz condition
\begin{equation}
A^\nu{}_{,\nu}=0.
\label{A}
\end{equation}

To $A^\nu$, satisfying Eqs. \eqref{em-wave} and \eqref{A}, we can apply conditions \ref{phi-O}, \ref{psi}. It would perhaps be more satisfactory if we formulated the boundary conditions in a way involving only the {\em field} $f_{\mu\nu}$. However, our actual conditions will be more suited for a straightforward generalization to Einstein's theory. The current $j^\alpha$ now satisfies the same regularity and boundness conditions as $\rho$  in the former case.

We formulate the boundary conditions as follows: {\em there exists a potential $A^\mu$  satisfying}
\begin{enumerate}[label=(\arabic*),leftmargin=0.89cm]
\setcounter{enumi}{\value{equation}}
\item $A^\mu=O(r^{-1})$ \label{Ar}
\item {\em and four functions $B_\mu=O(r^{-1})$ such that $A_{\rho,\sigma}=B_\rho k_\sigma+O(r^{-2})$, and } \label{AB}
\item $B_\rho k^\rho=O(r^{-2})$. \label{Bk}
\end{enumerate}
\setcounter{equation}{\value{enumi}}

It must be noted that there are many functions $A_\alpha$  which satisfy 
Maxwell's Eqs. \eqref{maxw} and conditions \ref{Ar}-\ref{Bk}, but all these potentials represent the same electromagnetic field.

From Eqs. \ref{AB}, \ref{Bk} we obtain the asymptotic form of the field\footnote{We shall sometimes write $\Phi\cong 0$  instead of $\Phi=O(r^{-2})$.}:
\begin{align}
f_{\mu\nu}&\cong k_\mu B_\nu-k_\nu B_\mu,& k_\rho B^\rho&\cong 0
\label{fmunu}
\end{align}
or, in vector notation,
\begin{align*}
  \vec{E}&\cong(\vec{B}\times \vec{n})\times\vec{n}, & \vec{H}&\cong\vec{B}\times\vec{n}, & B^\nu&=(B^0,\vec{B}), & k^\nu&=(1,\vec{n}).
\end{align*}
Eqs. \eqref{fmunu} represent a system of ``gauge-invariant'' boundary conditions. The electromagnetic field has asymptotically the form of a plane wave. For the energy-momentum density tensor
\[
4\pi T_{\mu\nu}=\frac{1}{4}\eta_{\mu\nu}f_{\rho\sigma}f^{\rho\sigma}-f_{\mu\rho}f_{\nu}{}^\rho
\]
we obtain the expression
\[
4\pi T_{\mu\nu}=-B_\rho B^\rho k_\mu k_\nu+O(r^{-3}).
\]
$k_\nu$ being a null vector, it follows from \ref{Bk} that $B_\rho B^\rho\leq 0$  for sufficiently large $r$.

The total charge $e$ contained in the field can be calculated by means 
of Gauss' law
\[
4\pi e=\oint_S f^{k0}n^k \,dS.
\]

Though $f^{k0}$  contains terms going as $1/r$, nevertheless $e$ is finite by 
virtue of \eqref{fmunu}.

\bigskip\medskip


\end{document}